\documentclass[twocolumn,superscriptaddress,aps,preprintnumbers,amsmath,amssymb,prl,nofootinbib]{revtex4}

\usepackage{graphicx}
\usepackage{epstopdf}
\usepackage{dcolumn}
\usepackage{bm}
\usepackage{hyperref}
\usepackage{color}

\usepackage{amsmath}

\begin{document}


\def\a{\alpha}
\def\b{\beta}
\def\c{\varepsilon}
\def\d{\delta}
\def\e{\epsilon}
\def\f{\phi}
\def\g{\gamma}
\def\h{\theta}
\def\k{\kappa}
\def\l{\lambda}
\def\m{\mu}
\def\n{\nu}
\def\p{\psi}
\def\q{\partial}
\def\r{\rho}
\def\s{\sigma}
\def\t{\tau}
\def\u{\upsilon}
\def\v{\varphi}
\def\w{\omega}
\def\x{\xi}
\def\y{\eta}
\def\z{\zeta}
\def\D{\Delta}
\def\G{\Gamma}
\def\H{\Theta}
\def\L{\Lambda}
\def\F{\Phi}
\def\P{\Psi}
\def\S{\Sigma}

\def\o{\over}
\def\beq{\begin{eqnarray}}
\def\eeq{\end{eqnarray}}
\newcommand{\gsim}{ \mathop{}_{\textstyle \sim}^{\textstyle >} }
\newcommand{\lsim}{ \mathop{}_{\textstyle \sim}^{\textstyle <} }
\newcommand{\vev}[1]{ \left\langle {#1} \right\rangle }
\newcommand{\bra}[1]{ \langle {#1} | }
\newcommand{\ket}[1]{ | {#1} \rangle }
\newcommand{\EV}{ {\rm eV} }
\newcommand{\KEV}{ {\rm keV} }
\newcommand{\MEV}{ {\rm MeV} }
\newcommand{\GEV}{ {\rm GeV} }
\newcommand{\TEV}{ {\rm TeV} }
\newcommand{\1}{\mbox{1}\hspace{-0.25em}\mbox{l}}
\def\diag{\mathop{\rm diag}\nolimits}
\def\Spin{\mathop{\rm Spin}}
\def\SO{\mathop{\rm SO}}
\def\O{\mathop{\rm O}}
\def\SU{\mathop{\rm SU}}
\def\U{\mathop{\rm U}}
\def\Sp{\mathop{\rm Sp}}
\def\SL{\mathop{\rm SL}}
\def\tr{\mathop{\rm tr}}

\def\dd{\mathrm{d}}
\def\ff{\mathrm{f}}
\def\BH{{\rm BH}}
\def\tot{{\rm tot}}
\def\ev{{\rm evap}}
\def\eq{{\rm eq}}
\def\SM{{\rm sm}}
\def\BSM{{\rm BSM}}
\def\Mpl{M_{\rm Pl}}
\def\GW{{\rm GW}}
\def\GeV{{\rm GeV}}
\def\mcP{\mathcal{P}}
\newcommand{\Red}[1]{\textcolor{red}{#1}}

\title{
Probing Beyond Standard Model via
Hawking Radiated Gravitational Waves
}

\author{Tomohiro Fujita}
\affiliation{Kavli Institute for the Physics and Mathematics of the Universe (Kavli IPMU), WPI, TODIAS, University of Tokyo, Kashiwa, 277-8583, Japan}

\begin{abstract}

We propose a novel technique to probe the beyond standard model (BSM) of particle physics. 
The mass spectrum of unknown BSM particles can be scanned
by observing gravitational waves (GWs) emitted by Hawking radiation of black holes. This is because information on 
the radiation of the BSM particles
is imprinted in the spectrum of the GWs.
We fully calculate the GW spectrum from evaporating black holes
taking into account the greybody factor.
As an observationally interesting application, 
we consider primordial black holes
which evaporate in the very early universe. 
In that case, since the frequencies of GWs are substantially redshifted,
the GWs emitted with the BSM energy scales become accessible
by observations.
\end{abstract}
\maketitle
\preprint{IPMU 14-0151}

%
\section{I.  Introduction}
%

Last year, Higgs particle is discovered~\cite{Aad:2012tfa} and all particles
in the standard model of particle physics are
eventually identified. 
However, many phenomena which cannot be explained within
the standard model have been found 
(e.g. dark matter, inflation, neutrino mass, etc). 
A number of hypothetical particles are introduced 
and supposed to be observed in the future.
Since those beyond standard model (BSM) particles 
are assumed to be very heavy and/or weakly coupling to 
the standard model particles, to detect them is not a easy task.
In fact, no evidence of a BSM particle is found in Large Hadron Collider, so far.
Therefore it is very important to consider a novel technique to probe
BSM particles.

In this paper, we propose a new way to scan the mass spectrum
of the BSM particles by using gravitational waves (GWs) which
are radiated by light black holes (BHs).
It is well known that light BHs lose their masses by emitting
particles through Hawking radiation and finally evaporate~\cite{Page:1976df, Hawking:1974sw}.
A BH emits only particles whose mass are smaller than 
Hawking temperature $T_\BH$,
\begin{equation}
M \lesssim T_\BH \equiv \Mpl^2 / M_\BH,
\end{equation}
where $\Mpl$ is the reduced Planck mass and $M_\BH$ is the mass of the BH.
$T_\BH$ increases as the BH loses its mass. Thus the BH begins to radiate
a heavy particle with a mass $M_\BSM$ when the Hawking temperature
reaches the mass, $T_\BH \simeq M_\BSM$.
Since Hawking temperature 
goes up to the Planck scale right before the evaporation of a BH, any particles whose masses are less than $\Mpl$ can be radiated by evaporating BHs. 

The mass spectrum of BSM particles is imprinted in the power spectrum of GWs from evaporating BHs. 
Roughly speaking,
this is because when a BH begins to emit a heavy
particle, the number of degrees of freedom (DOF) radiated by the BH changes
and the ratio between the energy going to GWs and the total radiative energy also changes. This drop of the energy  fraction causes 
a step like feature in the GW spectrum.
In eq.~\eqref{chain reaction}, the relationship between the BSM mass spectrum and the resultant GW spectrum is sketched,
\begin{equation}
\rho \left(M_\BSM \right) \rightarrow
g (T_\BH) \rightarrow
T_\BH (t) \rightarrow
\Omega_{\rm GW} (\nu_0).
\label{chain reaction}
\end{equation}
The BSM mass spectrum, $\rho \left(M_\BSM \right)$, determines
the DOF emitted by BHs as a function of
Hawking temperature, $g (T_\BH)$. The mass loss rate
of a BH is proportional to it, $\partial_t M_\BH(t) \propto g (T_\BH)$,
and we can solve the time evolution of the BH mass, $M_\BH(t),$ or 
equivalently, that of the Hawking temperature $T_\BH(t)$.
Then it is workable
to compute the resultant  spectrum
of the GWs, $\Omega_\GW(\nu_0)$, or any other particles. 
Note that the spectrum of photons or neutrinos can also
be candidates for observational probe but we focus
on the graviton case in this paper, because
the interaction with other particles 
is negligible.

The imprinted feature of the BSM physics in the GW spectrum
appears at the frequency which corresponds to the energy
scale of the BSM. Such a high frequency GW is perhaps undetectable.
However, if one identifies the BHs 
as primordial black holes (PBHs)~\cite{PBHreview}
which evaporate in the very early universe, the emitted GWs
are substantially redshifted and become accessible.

To obtain a proper spectrum form, we take into account
the greybody factor which is often ignored but 
significantly alters the spectrum. Moreover, since we do not know
the actual BSM theory, we assume that all the BSM particles live
at the GUT scale to demonstrate a readable spectrum.

The rest of paper is organized as follows.
In section 2, we briefly review Hawking radiation and greybody factor.
In section 3, the spectrum of GWs emitted by a BH without cosmic expansion
is calculated.
In section 4, the spectrum of GWs produced by PBHs 
is computed and its observability is discussed.
In section 5, we conclude.

%
\section{II. Hawking radiation}
%

In this section, let us briefly review Hawking radiation
and the greybody factor of gravitons.
The energy spectrum of a graviton emitted by the Hawking radiation
of a single BH per unit time is given by~\cite{Hawking:1974sw}
\begin{equation}
\frac{\dd E_\GW}{\dd t \dd \omega}
=
\frac{1}{2\pi}\frac{\omega}{e^{\omega/T_\BH}-1}
\times2 \Gamma (2GM_\BH\omega),
\label{Power}
\end{equation}
where $\omega$ is the energy of the graviton, $\Gamma$ denotes
the greybody factor (or the absorption coefficient) and the factor 2
in front of $\Gamma$ reflects the two polarization of graviton.
On the Black hole event horizon, particles are radiated
with the thermal distribution (blackbody) while not all of them  
reach a distant observer because of the gravitational potential
of the BH. The greybody factor,
$\Gamma$, represent the probability that a particle with
an energy $\omega$ travels to the infinite distance 
despite of the BH potential.
The greybody factor is obtained by solving
the equation of motion around the BH
of the particle in interest.
In the case of tensor perturbation around a Schwarzschild BH,
it is written as~\cite{Regge:1957td}
\begin{align}
\left[ \frac{\dd^2}{\dd x^2} + \omega^2 - V(r,l) \right]Q_l(r)=0,
\label{RW eq}\\ 
V(r,l)= \frac{r-1}{r}\left[ \frac{l(l+1)}{r^2}-\frac{3}{r^3}\right]=0,
\end{align}
where $r$ is the radial coordinate and  $x$ is Tortoise coordinate,
$x\equiv r + r\ln (r-1)$. Note that in this section, we set Schwarzschild radius as 1,
\begin{equation}
r_s \equiv 2GM_\BH=1.
\end{equation}
Eq.~\eqref{RW eq} is called ``Regge-Wheeler equation".
One can check the definition of $Q_l(r)$ in their paper~\cite{Regge:1957td},
but it is basically the $l$-mode of the graviton field 
whose polarization is odd.
The even mode has more complicated potential while its 
greybody factor is identical to the odd mode~\cite{Chandrasekhar:1985kt}.
Note that the label of the spherical harmonics should be $l\ge 2$ in the graviton case. 

\begin{figure}[tbp]
  \includegraphics[width=75mm]{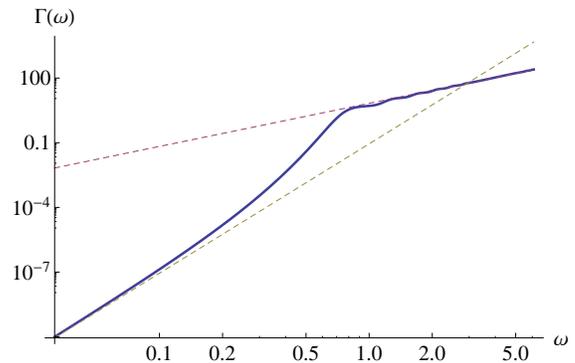}
 \caption
 {The blue solid line is the greybody factor of graviton introduced in eq.~\eqref{Power}  and numerically obtained based on eq.~\eqref{GRB eq}).
 The yellow and red dashed line represent the asymptotic behaviors
 in $\omega \ll 1$ (eq.~\ref{low energy behabior}) and $\omega \gg 1$ (eq.~\ref{high energy behabior}), respectively.}
 \label{GBF fig}
\end{figure}

Since the potential $V(r,l)$ vanish at $r\to 1$ and $\infty$, the asymptotic
solution of $Q_l$ is given by plane waves
\begin{align}
&Q_l \to e^{i\omega x} + R_l e^{-i \omega x} \quad (x\to-\infty;\ {\rm near\ horizon}),
\\
&Q_l \to T_l e^{i\omega x} \qquad\qquad(x\to +\infty;\ {\rm infinite\ distance}),
\end{align}
where $R_l$ and $T_l$ are the reflection, transmission coefficients, respectively.
Eq.~\eqref{RW eq} has the same form as  the Schr\"{o}dinger equation and hence we can use the analogy with the tunneling problem in quantum mechanics.
Then the greybody factor of graviton is given by
\begin{equation}
\Gamma(\omega) = \sum_{l=2} (2l+1) |T_l|^2.
\label{GRB eq}
\end{equation}
The analytic expressions of $|T_l|^2$ in
the low energy limit $\omega\ll 1$ and 
the greybody factor in the high energy limit $\omega \gg 1$ are known,
\begin{align}
&|T_l|^2 \xrightarrow{\omega\ll1} 4\pi \left(\frac{\omega}{2}\right)^{2l+2}
\left[\frac{\Gamma(l+3)\Gamma(l-1)}{\Gamma(2l+1)\Gamma(l+3/2)}\right]^2,
\label{low energy behabior}
\\
&\Gamma(\omega) \xrightarrow{\omega\gg1} \frac{27}{4}\omega^2.
\label{high energy behabior}
\end{align}
For general $\omega$, however, $T_l$ cannot be solved analytically
and a numerical calculation is needed.
We numerically obtain $\Gamma(\omega)$ and plot it in  fig.~\ref{GBF fig}. 
Our result is consistent with previous works~\cite{MacGibbon:1991tj,Page:1976df}.
Therefore by integrating eq.~\eqref{Power} with respect to time $t$,
the GW spectrum produced by a single BH can be obtained.

Before finishing this section, let us mention the effective DOF emitted by a BH. If one ignores greybody factor and
consider a BH as a black-body radiator, the Stefan-Boltzmann law yields
\begin{equation}
\frac{\dd E}{\dd t}
= \frac{\pi^2}{120}g T_\BH^4 A_\BH
=g \frac{\pi}{480} T_\BH^2
\label{Stefan}
\end{equation}
where $A_\BH\equiv 4\pi r_s^2$ is the area of the BH
and $g$ denotes the number of emitted DOF.
Comparing eqs.~\eqref{Power} and \eqref{Stefan},
Anantua et al.~\cite{Anantua:2008am} have introduced the following effective
DOF $g_s$ including the effect of greybody factors:
\begin{equation}
\frac{\dd E_s}{\dd t} =
 \int^\infty_0 \frac{\dd \omega}{2\pi}\frac{\omega \Gamma_s(\omega)}{e^{\omega/T_\BH}-1}
\equiv g_s \frac{\pi}{480}T_\BH^2,
\end{equation}
where $s$ is the spin of the particle in interest.
The values of $g_s$ are given by~\cite{MacGibbon:1991tj}
\begin{align}
&g_{s=0}\approx7.26,\quad
g_{s=1/2}^{\rm uncharged}\approx4.00,\quad
g_{s=1/2}^{\rm charged}\approx3.86,\quad
\notag\\
&g_{s=1}\approx1.63,\quad
g_{s=2}\approx0.185\,,
\end{align}
where  ``(un)charged" denotes the electric charge of the emitted spinor.
One obtains the total DOF in the standard model
as~\cite{comment1}
\begin{equation}
g_{\rm SM} \approx 4\times10^2\,.
\end{equation}
Therefore after all the standard model particles
are begun to radiate, less than $0.1\%$ of the total emitted energy
is radiated as gravitons. Note that this result is different by 
a order of magnitude from the naive estimation by the effective DOF
in thermal equilibrium, $2/106.75\approx 2\%$.

%
\section{III. GW spectrum from evaporating BH}
%

In this section, we calculate the GW spectrum
produced by a single BH without cosmic expansion.
For simplicity, we consider that the total effective
DOF changes instantly and only once at a BSM mass scale,
\begin{equation}
g_{\rm tot}(T_\BH)
= \left\{ 
\begin{array}{ll}
g_1 &\ ( T_\BH < M_{\BSM} ) \\
g_2  &\ ( T_\BH > M_{\BSM} )
\end{array}\right..
\label{g change}
\end{equation}
Then solving the evolution equation of a BH mass,
\begin{equation}
\frac{\dd M_\BH}{\dd t}
 = -g_{\rm tot}(T_\BH) \frac{\pi}{480}T_\BH^2,
\end{equation}
one can obtain the time evolution of $T_\BH$ as~\cite{Fujita:2013bka}
\begin{equation}
T_\BH (t) =
\left\{ 
\begin{array}{ll}
T_0(1-\frac{t}{\tau_1})^{-1/3} &\, (0< t < t_c) \\
M_\BSM(1-\frac{t-t_c}{\tau_2})^{-1/3}  &\, ( t_c<t< \tau_\tot)
\end{array}\right.,
\label{T evolution}
\end{equation}
where $T_0 \equiv T_\BH(0)$ is the initial Hawking temperature,
$\tau_1 \equiv 160\Mpl^2/\pi g_1 T_0^3$ is the lifetime of the BH
if $g_\tot=g_1$ regardless of $T_\BH$, $t_c \equiv \tau_1 ( 1- T_0^3/M_\BSM^3)$ is the time
when $g_\tot$ changes, $\tau_2 \equiv 160 \Mpl^2/ \pi g_2 M_\BSM^3$
is the lifetime after $t=t_c$ and $\tau_\tot \equiv t_c+\tau_2$
is the total lifetime of the BH. 

\begin{figure}[tbp]
  \includegraphics[width=75mm]{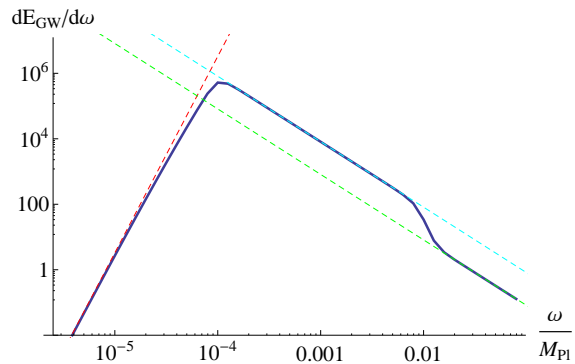}
 \caption
 {The blue solid line is the GW spectrum produced by a single BH (see eq.~\eqref{GW   spe without expansion}).
  We set parameters as $T_0=10^{-5}\Mpl, M_\BSM=10^{-3}\Mpl$ (GUT scale),
  $ g_1=g_{\rm SM}$ and $g_2=10g_{\rm SM}$.
 The step like feature appears at $\omega \approx 10 M_\BSM$
 and the amplitude drops there by the factor of $g_1/g_2$.
 The red, cyan and green dashed lines represent the asymptotic behaviors
given by eqs.~\eqref{low approx} and \eqref{high approx}.}
 \label{dEdw fig}
\end{figure}

Substituting eq.~\eqref{T evolution} into eq.~\eqref{Power},
we obtain the time derivative of the graviton spectrum, $\dd E_\GW/\dd t \dd \omega$,
as a function of time. 
Nonetheless, it is important to notice that if the BSM scale $M_\BSM$
is much higher than the experimentally accessible scale, we cannot resolve the time variability of $\dd E_\GW/\dd t \dd \omega$.
For example, provided $M_\BSM \gg T_0 = 10^{-5}\Mpl$ and $g_1 = g_{\rm SM}$,
the BH lifetime is $\tau_\tot \approx 10^{14}\Mpl^{-1} \approx 3\times 10^{-29}$sec.
Therefore, in practice, the BH evaporates instantaneously and the observed
spectrum is the time integral of eq.~\eqref{Power}.
Then we find
\begin{align}
\frac{\dd E_\GW}{\dd \omega} 
&=\frac{\omega}{\pi}
\int_0^{\tau_\tot}\dd t \frac{\Gamma(\omega/4\pi T_\BH(t))}{e^{\omega/T_\BH(t)}-1},
\label{GW spe without expansion}
\\
&= \frac{480}{\pi^2 g_1} \frac{\Mpl^2}{\omega^2}\left[
\int^{\omega/T_0}_{\omega/M_\BSM}\dd X \frac{X^2 \Gamma(X/4\pi)}{e^X -1}
\right. \notag\\
&\qquad\ \ \  + \left.
\frac{g_1}{g_2}\int^{\omega/M_\BSM}_0 \dd X \frac{X^2\Gamma(X/4\pi)}{e^X-1}\right],
\label{X form}
\end{align}
where we define $X\equiv \omega/ T_\BH(t)$.
The integrand in eq.~\eqref{X form} has a peak at $X\approx 10$ since
the greybody factor $\Gamma(x)$ is suppressed for $x \lesssim 0.8$ 
(see fig.~\ref{GBF fig}). Therefore gravitons with energy $\omega$ are mostly emitted when $\omega \approx 10T_\BH(t)$.
This GW spectrum is numerically evaluated and plotted in fig.~\ref{dEdw fig}.
%
%

Let us derive asymptotic expressions of eq.~\eqref{X form}.
For $\omega \ll T_0$, the first line in eq.~\eqref{X form}
has the main contribution.
Using eq.~\eqref{low energy behabior}
with $l=2$, namely $\Gamma(x)\sim 4x^6/45$, and the Taylor expansion in the denominator,
one can show
\begin{equation}
\frac{\dd E_\GW}{\dd \omega} \xrightarrow{\omega\ll T_0}
\frac{\pi^{-8}}{768g_1}\frac{\omega^6 \Mpl^2}{T_0^8}.
\label{low approx}
\end{equation}
On the other hand, for $\omega \gg T_0$,
$\omega$ is greater than $T_\BH(t)$ at the beginning while
$T_\BH(t)$ finally becomes much larger than $\omega$. 
Thus the integration interval can be approximated by $X=[0,\infty]$.
The numerical evaluation yields
\begin{equation}
\int^{\infty}_{0}\dd X \frac{X^2 \Gamma(X/4\pi)}{e^X -1}
\approx 0.07\,,
\label{numerical integral}
\end{equation}
and one finds
\begin{equation}
\frac{\dd E_\GW}{\dd\omega}\xrightarrow{\omega\gg T_0}
\frac{3.4}{g_{1,2}}\frac{\Mpl^2}{\omega^2},
\label{high approx}
\end{equation}
where $g_{1,2}$ is $g_1$ for $\omega \lesssim 10M_\BSM$
and $g_2$ for $\omega \gtrsim 10M_\BSM$.
These approximated spectra, which are plotted in fig.~\ref{dEdw fig} as dashed lines, clearly explain that the step like feature
appears at $\omega \approx 10M_\BSM$ and the amplitude drops there by the factor
of $g_1/g_2$. The reason of the drop of the amplitude can be understood that the energy ratio going to gravitons decreases as the total DOF of the Hawking radiation increases.

The GW spectrum, fig.~\ref{dEdw fig}, can be realized
if a single BH evaporates in our neighborhood in which 
the cosmic expansion is negligible.
Although $T_0$ should be taken much lower in that case,
it does not affect the step like feature.
Thus if we could observe such spectrum, it is possible to know
the mass scale and the DOF, namely the mass spectrum,
of BSM particles.
Unfortunately, however, it is difficult to observe the step
in this case because its frequency is around $M_\BSM$ and 
is probably too high to be detected even in the future.

In the next section,
we consider primordial black holes (PBHs) which evaporate
in very early universe.
The frequency of a graviton which was emitted by a PBH
gets substantially redshifted before coming to the earth
and hence its frequency can be low enough to be observed. 

%
\section{IV. GW spectrum from PBH}
%

In this section, we calculate the GW spectrum
produced by PBHs. The GW spectrum from PBHs
has been computed in previous works~\cite{Dolgov:2011cq, Anantua:2008am}
but neither the greybody factor nor the change of the DOF are
taken into account (however the latter is discussed in ref.~\cite{Fujita:2014hha}).
In the case of PBHs, two additional effect should be considered; cosmic expansion and the number density of  PBHs.

PBHs are formed at
\begin{equation}
t_{\rm form} \simeq (8\pi \gamma T_0)^{-1}
\end{equation}
where the initial mass of a PBH is given by
$M_0 = 4\pi \gamma \rho/3H^3(t_{\rm form})$ and  $T_0 \equiv \Mpl^2/M_0$. Provided that PBHs are formed at the radiation dominant era,
the PBH energy fraction increases, $\Omega_\BH \equiv \rho_\BH/3\Mpl^2H^2\propto a$. Therefore if the initial energy fraction, $\beta \equiv \Omega_\BH(t_{\rm form})$,
is large enough, $\beta \gtrsim \sqrt{g_1}\Mpl/36\sqrt{\gamma} M_0$, PBHs dominate the universe before their evaporation at $t_{\rm evap}\equiv t_{\rm form}+\tau_\tot \simeq t_c$.
In that case, from the onset of the PBH domination until the evaporation,
the universe is in matter dominant era and
the total energy density at the evaporation is given by
\begin{equation}
\rho_{\rm evap} \simeq \frac{4\Mpl^2}{3t_c^2}.
\end{equation}
Ignoring the change of the DOF in the thermal bath, one finds the scale
factor at the evaporation is
\begin{equation}
a_{\rm evap} \simeq a_\eq \left(\frac{\rho_\eq}{\rho_{\rm evap}}\right)^{1/4}
\simeq \left( \frac{3a_\eq t_c^2 \rho_{\rm now}}{4\Mpl^2} \right)^{1/4},
\end{equation}
where the subscript ``eq" denotes the time of matter-radiation equality
and $\rho_{\rm now}$ is the energy density at present.
Using the scaling, $a\propto t^{1/2}$ during radiation dominant era
and $a\propto t^{2/3}$ during matter dominant era, one can obtain
the scale factor $a(t)$ from the PBH formation until the evaporation.

\begin{figure}[tbp]
  \includegraphics[width=75mm]{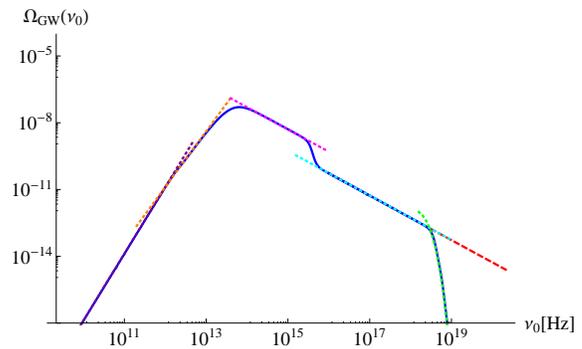}
 \caption
 {The blue solid line is $\Omega_\GW(\nu_0)$ produced by evaporating PBHs (see eq.~\eqref{PBH GW equation}).
  The parameters are same as fig.~\ref{dEdw fig} with $\gamma =0.2$ and $\beta = 10^{-4}$.
 The step like feature appears at $\nu_0 \approx 10^{15} $Hz
 and the amplitude drops there by the factor of $g_1/g_2$.
 The cutoff at $\nu_0 \approx 10^{18}$Hz is introduced by hand
 because the contribution from $T_\BH \ge \Mpl$ is not reliable.
 The red dashed line show the case without the cutoff.
 The colorful dotted lines represent approximated  spectra derived in appendix 
and confirm the validity of the numerical calculation.}
 \label{PBHGW fig}
\end{figure}

Remembering $\omega(t) = 2\pi \nu_0/a(t)$ where $\nu_0$ is the comoving
frequency, we find that $\Omega_\GW \equiv \rho_{\rm now}^{-1}\dd \rho_\GW/\dd \ln \nu_0$ of the Hawking radiated gravitons at present is written by
\begin{equation}
\Omega_\GW(\nu_0)= \frac{4\pi\nu_0^2}{\rho_{\rm now}}\int^{t_{\rm evap}}_{t_{\rm form}}\dd t\,
\frac{a^{2} n_\BH \Gamma(\nu_0/2aT_\BH)}{e^{2\pi\nu_0/aT_\BH}-1},
\label{PBH GW equation}
\end{equation}
where $n_\BH$ is the PBH number density with the initial value,
$n_\BH(t_{\rm form})=3\beta \Mpl^2/ 4M_0 t_{\rm form}^2$.
We numerically evaluate this equation and plot it in fig.~\ref{PBHGW fig}.
%
%
Again one can see the step like feature in the GW spectrum.
Furthermore, in the case of fig.~\ref{PBHGW fig}, 
the frequency of the step is redshifted by the factor of $a_{\rm evap} \approx 10^{-24}$,
and given by 
\begin{equation}
\frac{2\pi\nu_{\rm step}}{a_{\rm evap}}\approx 10M_\BSM
\ \Longleftrightarrow\ 
\nu_{\rm step} \approx 10^{15} {\rm Hz}.
\end{equation}
Thus the frequency of the step is now accessible
(remember $(4-8)\times10^{14}$Hz is the frequency of
visible light).
In fact, the GW detector built by the group in university of Birmingham
has sensitivity at $\nu_0 \approx 10^{15}$Hz~\cite{Cruise:2012zz}.
Although the sensitivity is not enough at present, it is expected to 
increase significantly in the future~\cite{Li:2008qr}.

It should be noted that the step frequency $\nu_{\rm step}$ depend on the initial mass of the PBHs. Here we consider that the PBHs are formed
right after inflation due to the preheating~\cite{Suyama:2004mz} 
while PBHs can be formed by many other processes~\cite{PBHreview}.
Then $M_0\simeq10^{5}\Mpl$ is obtained 
from the Hubble parameter right after inflation,
$H_f \simeq 10^{13}$GeV, which is favored based on the 
BICEP2 result~\cite{Ade:2014xna}. 

The Hawking radiation, eq.~\eqref{Power}, is derived based on the quasi-classical treatment and is no longer reliable for $\omega\gtrsim \Mpl$.
Therefore we introduce the cutoff in the integral range 
of eq.\eqref{PBH GW equation} by replacing $t_{\rm evap}$ by 
$t_p\equiv t_{\rm evap}-160/\pi g_2 \Mpl$ at which the Hawking temperature
reaches $\Mpl$.
Because of this artificial cutoff, $\Omega_\GW(\nu_0)$ in fig.\ref{PBHGW fig} rapidly falls at $\nu_{\rm cut} \approx 10^{18}$Hz while the red dashed line shows the case without the cutoff.

%
\section{V. Summary and Discussion}
%

In this paper, we demonstrate that if the DOF of Hawking radiation 
increases at a BSM scale, a step like feature 
is imprinted in the GW spectrum
produced by evaporating BHs. 
Since the step height and the frequency of 
the feature indicate the number of additional DOF
and the energy scale of BSM particles, respectively,
we can scan the mass spectrum of the actual BSM theory
by observing the GW spectrum.
We assume that all BSM particles live at $10^{-3}\Mpl$
for simplicity, set the initial mass of the PBH as $10^{5}\Mpl$
inspired by the BICEP2 result,
and calculate the GW spectrum from the PBHs (see fig.~\ref{PBHGW fig}).
It is found that the frequency of the spectrum feature
is substantially redshifted due to cosmic expansion 
and enters the observable range.

In reality, the BSM mass spectrum may be distributed over
many different energy scales. In that case, a lot of steps
appear in the GW spectrum while our methodology is still useful.
Note that BHs can radiate even ``dark particles"
which couple to the standard model sector very weakly. 
Therefore our technique is sensitive to these dark particles
and can be complementary to particle accelerators or direct 
detection experiments.

%
\section{Acknowledgements}
%
We would like to thank Teruaki Suyama for useful discussions.
This work is supported by 
World Premier International Research Center Initiative
(WPI Initiative), MEXT, Japan.
The author  acknowledges 
JSPS Research Fellowship for Young Scientists, No.248160.

%
\appendix
\section{APPENDIX: approximated analytic spectra}
%

In this appendix, we derive the approximated analytic spectra
plotted in fig.~\ref{PBHGW fig} as the dotted lines
in order to cross-check our numerical result.
Calculational procedures are almost same as eqs.~\eqref{low approx}
and \eqref{high approx}.

%
\subsection{1. $\omega(t_{\rm form}) \ll 10T_0$}
%

For this range of $\omega$, the peak contribution from $\omega \simeq 10T_\BH$
is never gained. Using the low energy approximations,
$\Gamma(x) \simeq 4x^6/45$ and $e^x \simeq 1+x$,
one finds
\begin{equation}
\Omega_\GW(\nu_0) \simeq \frac{\nu_0^7}{360\rho_{\rm now}}
\int^{t_{\rm evap}}_{t_{\rm form}} \dd t \frac{n_\BH}{a^3 T_\BH^5}.
\end{equation}
Since the biggest contribution comes from $t\sim t_{\rm form}$,
$T_\BH$ is approximated by $T_0$ and the above equation reads
\begin{equation}
\Omega_\GW(\nu_0) \simeq \frac{\nu_0^7 n_{\BH,0}}{360\rho_{\rm now} a_{\rm form}^3T_0^5}
\int^{t_{\rm evap}}_{t_{\rm form}} \dd t \left(\frac{t_{\rm form}}{t}\right)^3,
\end{equation}
where $n_{\BH,0}\equiv n_\BH(t_{\rm form})$ which can be 
rewritten as 
$n_{\BH,0}=3\Mpl^3\beta/ 4t_{\rm form}^2M_0$. Then we obtain
\begin{equation}
\Omega_\GW(\nu_0) \simeq \frac{\pi\beta\gamma}{120}\frac{a_{\rm form}^{-3}\nu_0^7}{\rho_{\rm now}T_0^3}.
\end{equation}
In fig.~\ref{PBHGW fig}, this region of $\nu_0$ is too small to be plotted.

%
\subsection{2. $\omega(t_{\rm form}) \gg 10T_0 \gg \omega(t_c)$}
%

In this range, $\omega$ becomes comparable to $10T_\BH$
because $\omega\ (\propto a^{-1})$ decreases while
$T_\BH$ remains almost constant at $T_0$.
Approximating $T_\BH(t)$ by $T_0$ and
ignoring the contribution from $t>t_c$, one can show
\begin{align}
\Omega_\GW(\nu_0) &\simeq
\frac{4\pi\nu_0^2}{\rho_{\rm now}}a_{\rm form}^3 n_\BH(t_{\rm form})
\notag\\
&\times\left[
\frac{4\pi\nu_0 t_{\rm form}}{a_{\rm form}^2 T_0}
\int^{\frac{2\pi \nu_0}{a_{\rm form}T_0}}_{\frac{2\pi \nu_0}{a_{\rm dom}T_0}}
\dd Y \frac{Y^{-2}\Gamma(Y/4\pi)}{e^Y-1}\right.
\notag\\
+
&\left. 
3t_c\sqrt{\frac{\pi\nu_0}{2a_{\rm evap}^3T_0}} 
\int^{\frac{2\pi \nu_0}{a_{\rm dom}T_0}}_{\frac{2\pi \nu_0}{a_{\rm c}T_0}}
\dd Z \frac{Z^{-3/2}\Gamma(Z/4\pi)}{e^Z-1}
\right],
\label{middle region}
\end{align}
where the subscript ``dom" denotes the time when the PBHs dominate
the universe. Two integrand have the peak at $Y\approx 9$ and
$Z \approx 8$, and the numerical integral with the approximated
interval $[0,\infty]$ yield,
\begin{align}
\int^\infty_0 \dd Y \frac{Y^{-2}\Gamma(Y/4\pi)}{e^Y-1} \approx 1.4\times10^{-5},
\\
\int^\infty_0 \dd Z \frac{Z^{-3/2}\Gamma(Z/4\pi)}{e^Z-1} \approx 3.8\times10^{-5}.
\end{align}
Therefore eq.~\eqref{middle region} reads
\begin{align}
&\Omega_\GW(\nu_0) \simeq 4\times10^{-2} \frac{a_{\rm form}\gamma \beta T_0}{\rho_{\rm now}}\nu_0^3
\notag\\
&~~~~~~~~~~~~~~~~~~~~~~~~~~~(a_{\rm dom}\gg \frac{2\pi \nu_0}{9T_0}\gg a_{\rm form}),
\\
&\Omega_\GW(\nu_0) \simeq 2\times10^{-3} \frac{a_{\rm form}^3 n_{\BH,0}t_c}{\rho_{\rm now}a_{\rm evap}^{3/2} T_0^{1/2}}\nu_0^{5/2}
\notag\\
&~~~~~~~~~~~~~~~~~~~~~~~~~~~  (a(t_c)\gg \frac{2\pi \nu_0}{8T_0}\gg a_{\rm dom}).
\end{align}
%
They are shown as the purple and orange dotted lines in fig.~\ref{PBHGW fig}.

%
\subsection{3. $\omega(t_c) \gg 10T_0 $}
%

For $t\gtrsim 0.1 t_c$, the time variation of $T_\BH$ is significant
while the cosmic expansion is negligible. One can show
\begin{align}
\Omega_\GW(\nu_0) &\simeq
\frac{4\pi\nu_0^2}{\rho_{\rm now}}a_{\rm form}^3 n_{\BH,0} a_{\rm evap}^{-1}
\notag\\
&\times\left[
3\tau_1 \left(\frac{a_{\rm evap}T_0}{2\pi\nu_0}\right)^3
\int^{\frac{2\pi \nu_0}{a_{\rm evap}T_0}}_{\frac{2\pi \nu_0}{a_{\rm evap}M_\BSM}}
\dd X \frac{X^{2}\Gamma(X/4\pi)}{e^X-1}\right.
\notag\\
+
&\left. 
3\tau_2 \left(\frac{a_{\rm evap}M_\BSM}{2\pi\nu_0}\right)^3
\int^{\frac{2\pi \nu_0}{a_{\rm evap}M_\BSM}}_0
\dd X  \frac{X^{2}\Gamma(X/4\pi)}{e^X-1}
\right].
\label{high region}
\end{align}
Here, the time integral is split into the two parts
because of the time dependence of $T_\BH(t)$ (see eq.~\eqref{T evolution}).
Using eq.~\eqref{numerical integral}, we obtain
\begin{align}
&\Omega_\GW(\nu_0) \simeq 10^{-2} a_{\rm form}^3 a_{\rm evap}^2
\frac{T_0^3 n_{\BH,0} \tau_1}{\rho_{\rm now}}\nu_0^{-1}
\notag \\
&~~~~~~~~~~~~~~~~~~~~~~~~~~~~~(T_0\ll \frac{2\pi \nu_0}{10a_{\rm evap}}< M_\BSM),
\\
&\Omega_\GW(\nu_0) \simeq 10^{-2} a_{\rm form}^3 a_{\rm evap}^2
\frac{M_\BSM^3 n_{\BH,0} \tau_2}{\rho_{\rm now}}\nu_0^{-1}
\notag \\
&~~~~~~~~~~~~~~~~~~~~~~~~~~~~~(M_\BSM \ll \frac{2\pi \nu_0}{10a_{\rm evap}}< \Mpl).
\end{align}
They are the magenta and cyan dotted lines in fig.~\ref{PBHGW fig}.
The ratio between two spectra is 
\begin{equation}
\frac{\Omega_\GW(\nu_{\rm high})}{\Omega_\GW(\nu_{\rm low})}
\simeq \frac{M_\BSM^3\tau_2}{T_0^3 \tau_1} =\frac{g_1}{g_2},
\end{equation}
and it reflects the change of the DOF.

%
\subsection{4. $\omega(t_{\rm evap}) \ge \Mpl $}
%

For the sake of completeness, let us obtain the analytic
expression for $\nu_0 > \nu_{\rm cut}$.
Considering the contribution from $t\ge t_p$,
one can find
\begin{align}
\Omega_\GW(\nu_0) &\simeq
\frac{3}{2\pi^2}a_{\rm form}^3  a_{\rm evap}^{2}
\frac{n_{\BH,0}\tau_2 M_\BSM^3}{\rho_{\rm now}\nu_0}
\notag\\
&\times
\int^\infty_{\frac{2\pi \nu_0}{a_{\rm evap}\Mpl}}
\dd X \frac{X^{2}\Gamma(X/4\pi)}{e^X-1}.
\end{align}
In this region, one see $X\gg1$ and the high energy
approximation, $\Gamma(x) \simeq 27x^2/4$, can be used.
Then it reads
\begin{equation}
\Omega_\GW \simeq \frac{1620}{\pi}\frac{a_{\rm form}^3}{a_{\rm evap}^2}
\frac{n_{\BH,0}\nu_0^3}{g_2 \rho_{\rm now}\Mpl^2}
\exp\left[-\frac{2\pi\nu_0}{a_{\rm evap}\Mpl}\right].
\end{equation}
It is shown as the green dotted line in fig.~\ref{PBHGW fig}.
Note that in this region, a graviton physical energy at the emission
exceeds the Planck scale and no reliable treatment is  established.



\begin{thebibliography}{99}
\bibitem{comment1}
Following ref.~\cite{MacGibbon:1991tj}, one finds $g_{\rm SM}\approx 416.8$.
Nevertheless, since the electric charges of the quarks are approximated by $\pm e$
in ref.~\cite{MacGibbon:1991tj}, this $g_{\rm SM}$ may be underestimated
by a few \% level.
\end{thebibliography}
\end{document}